\documentclass[10pt,conference]{IEEEtran}

\usepackage{colortbl}
\usepackage{makecell}
\usepackage{hyperref}
\usepackage{boldline}
\usepackage{color}
\usepackage{bigstrut}
\usepackage[ruled]{algorithm2e}
\usepackage{amsmath,amsfonts}
\usepackage{amsmath}
\usepackage{array}
\usepackage{algorithmic}
\usepackage{balance}
\usepackage{blindtext}
\usepackage{booktabs}
\usepackage{hyperref}
\usepackage[noabbrev]{cleveref}
\usepackage{comment}
\usepackage{enumitem}
\usepackage{eqparbox}
\usepackage{fancybox}
\usepackage{fancyvrb}
\usepackage{framed}
\usepackage{graphicx}
\usepackage{ifthen}
\usepackage{listings}
\usepackage{mathrsfs}
\usepackage{mdwmath}
\usepackage{multirow}
\usepackage{pifont}
\usepackage{textcomp}
\usepackage{url}
\usepackage{xspace}
\usepackage[normalem]{ulem}
\usepackage[framemethod=TikZ]{mdframed}
\usepackage{caption}
\usepackage{subcaption}
\usepackage{tabularx}
\usepackage{tcolorbox}
\tcbuselibrary{listings,skins}
\usepackage{mathtools}
\usepackage{blindtext}
\usepackage{lstautogobble}
\usepackage[numbers,sort&compress]{natbib}

\DeclareGraphicsExtensions{.pdf,.jpeg,.png}
\graphicspath{{figures/}}

\usetikzlibrary{shadows}
\usepackage{graphics}
\newmdenv[
    tikzsetting= {fill=blueish},
    skipabove=0.33em,
    skipbelow=0.33em,
    linewidth=1pt,
    innerleftmargin=4pt,
    innerrightmargin=4pt,
    innertopmargin=2pt,
    innerbottommargin=2pt,
    linecolor=gray95,
    roundcorner=2pt, 
    shadow=true,
    shadowsize=4pt,
    shadowcolor=gray95
]{questionbox}

\newmdenv[
    tikzsetting= {fill=greenish},
    skipabove=0.33em,
    skipbelow=0.33em,
    linewidth=1pt,
    innerleftmargin=4pt,
    innerrightmargin=4pt,
    innertopmargin=2pt,
    innerbottommargin=2pt,
    linecolor=gray95,
    roundcorner=2pt, 
    shadow=true,
    shadowsize=4pt,
    shadowcolor=gray95
]{answerbox}

\newmdenv[
    skipabove=0.33em,
    skipbelow=0.33em,
    innerleftmargin=4pt,
    innerrightmargin=4pt,
    innertopmargin=2pt,
    innerbottommargin=2pt,
]{lessonbox}

\usepackage{tikz}

\definecolor{javared}{rgb}{0.6,0,0} %
\definecolor{javagreen}{rgb}{0.25,0.5,0.35} %
\definecolor{javapurple}{rgb}{0.5,0,0.35} %
\definecolor{javadocblue}{rgb}{0.25,0.35,0.75} %

\definecolor{blueish}{RGB}{250, 250, 255}
\definecolor{greenish}{RGB}{250, 255, 250}
\definecolor{redish}{RGB}{255, 200, 200}

\definecolor{highlight}{RGB}{175, 255, 100}
\definecolor{gray01}{gray}{.98}
\definecolor{gray05}{gray}{0.95}
\definecolor{gray08}{gray}{0.92}
\definecolor{gray10}{gray}{0.90}
\definecolor{gray12}{gray}{0.88}
\definecolor{gray15}{gray}{0.85}
\definecolor{gray18}{gray}{0.82}
\definecolor{gray20}{gray}{0.80}
\definecolor{gray25}{gray}{0.75}
\definecolor{gray30}{gray}{0.70}
\definecolor{gray35}{gray}{0.65}
\definecolor{gray40}{gray}{0.60}
\definecolor{gray45}{gray}{0.55}
\definecolor{gray50}{gray}{0.50}
\definecolor{gray55}{gray}{0.45}
\definecolor{gray60}{gray}{0.40}
\definecolor{gray65}{gray}{0.35}
\definecolor{gray70}{gray}{0.30}
\definecolor{gray75}{gray}{0.25}
\definecolor{gray80}{gray}{0.20}
\definecolor{gray85}{gray}{0.15}
\definecolor{gray90}{gray}{0.10}
\definecolor{gray95}{gray}{0.05}
\definecolor{rowgray}{RGB}{224, 224, 224}

\newcommand{\cmark}{\ding{51}}%
\newcommand{\xmark}{\ding{55}}%
\newcommand{\mycmark}{\color{teal} \cmark \color{black}}
\newcommand{\mycmarkreverse}{\color{brown} \cmark \color{black}}
\newcommand{\myxmark}{\color{brown} \xmark \color{black}}
\newcommand{\myxmarkreverse}{\color{teal} \xmark \color{black}}
\newcommand{\colorblue}{\cellcolor[rgb]{0.83,0.92,1}}

\newcommand{\colorgray}{\cellcolor[rgb]{0.906, 0.902, 0.902}}

\newcommand{\dataset}{\textsc{PrimeVul}\xspace}

\newcommand{\vd}{\textsc{vd}\xspace}

\newcounter{findingCounter}
\newenvironment{finding}{
\begin{tcolorbox}[colback=blue!5!white,colframe=blue!5!white,arc=0mm,grow to left by=0mm,left=0mm,grow to right by=0mm,left=1.5mm,right=1.5mm,top=1.5mm,bottom=1.5mm]
\textbf{Result-\arabic{findingCounter}\stepcounter{findingCounter}:}
}
{
\end{tcolorbox}
}

\begin{document}

\title{Vulnerability Detection with Code Language Models: How Far Are We?}

\author{Yangruibo Ding$^\dagger$, Yanjun Fu$^\diamondsuit$, Omniyyah Ibrahim$^\natural$, Chawin Sitawarin$^\ddagger$, Xinyun Chen$^\nabla$ \\
Basel Alomair $^{\natural,*}$, David Wagner $^\ddagger$, Baishakhi Ray$^\dagger$, Yizheng Chen$^\diamondsuit$ \\
\\
$^\dagger$\textit{Columbia University}\\
$^*$\textit{University of Washington}\\
$^\natural$\textit{King Abdulaziz City for Science and Technology} \\
$^\nabla$\textit{Google DeepMind} \\
$^\ddagger$\textit{UC Berkeley} \\
$^\diamondsuit$\textit{University of Maryland} 
}

\maketitle
\thispagestyle{plain}
\pagestyle{plain}

\begin{abstract}
In the context of the rising interest in code language models (code LMs) and vulnerability detection, we study the effectiveness of code LMs for detecting vulnerabilities.
Our analysis reveals significant shortcomings in existing vulnerability datasets, including poor data quality, low label accuracy, and high duplication rates, leading to unreliable model performance in realistic vulnerability detection scenarios. 
Additionally, the evaluation methods used with these datasets are not representative of real-world vulnerability detection.

To address these challenges, we introduce \dataset, a new dataset for training and evaluating code LMs for vulnerability detection. \dataset incorporates a novel set of data labeling techniques that achieve comparable label accuracy to human-verified benchmarks while significantly expanding the dataset.
It also implements a rigorous data de-duplication and chronological data splitting strategy to mitigate data leakage issues, alongside introducing more realistic evaluation metrics and settings. This comprehensive approach aims to provide a more accurate assessment of code LMs' performance in real-world conditions.

Evaluating code LMs on \dataset reveals that existing benchmarks significantly overestimate the performance of these models.
For instance, a state-of-the-art 7B model scored 68.26\% F1 on BigVul but only 3.09\% F1 on \dataset. Attempts to improve performance through advanced training techniques and larger models like GPT-3.5 and GPT-4 were unsuccessful, with results akin to random guessing in the most stringent settings. These findings underscore the considerable gap between current capabilities and the practical requirements for deploying code LMs in security roles, highlighting the need for more innovative research in this domain.
\end{abstract}

\section{Introduction}

In the evolving landscape of software development, code language models (code LMs) have become pivotal in automating various software engineering tasks, fundamentally altering developers' coding approaches~\cite{hou2024large}. Leading examples like GitHub Copilot~\cite{github-2021-copilot} and Amazon CodeWhisperer~\cite{amazon-2023-codewhisperer} have already integrated into real-world software development, significantly assisting developers in their daily work. Consequently, LM-based vulnerability detection (\vd) has gained traction, with researchers utilizing code LMs' expanding capabilities to autonomously identify security vulnerabilities in codebases~\cite{chakraborty2021deep, chen2023diversevul, steenhoek2023empirical, fu2022linevul}.

In this paper, we aim to evaluate whether code LMs are able to detect security vulnerabilities in real code, in settings representative of that needed for real-world use. 
This exploration is anchored in the belief that while code LMs possess remarkable potential, realizing their full capability to enhance software security necessitates a rigorous validation of their {\em training} and {\em evaluation} frameworks against the challenges of real-world software development.
In particular, we scrutinize the datasets used to train code LMs and the benchmarks and metrics used to evaluate their effectiveness at vulnerability detection.

\smallskip
\noindent
\textbf{Limitation of existing datasets and benchmarks.}
First, we meticulously analyze existing \vd benchmarks~\cite{lu2021codexglue, chen2023diversevul, fan2020bigvul, bhandari2021cvefixes, nikitopoulos2021crossvul}, examining data collection methods, label accuracy, and prevalence of data duplication.
Our investigation reveals critical data quality problems, that impact their effectiveness for training and their suitability for evaluating code LMs.

\paragraph*{Noisy labels}
In \vd literature, researchers typically label datasets either automatically or manually. Most large datasets~\cite{chakraborty2021deep, chen2023diversevul, fan2020bigvul} use automatic labeling because manual labeling is too expensive.
However, automatic labeling can introduce significant label noise. For instance, datasets like BigVul~\cite{fan2020bigvul} curate hundreds of thousands of functions from the real world and rely on vulnerability-fixing commits for labeling. However, they suffer from a flawed assumption that each function modified by such a commit corresponds to a (separate) vulnerability. In practice, vulnerability-fixing commits often fix one vulnerability but also make other changes to surrounding code, and existing automatic labeling methods wrongly label that surrounding code as vulnerable. In contrast, manual labeling offers higher accuracy, but its cost means it can only be applied to smaller datasets.  For instance, the most accurate prior dataset, SVEN~\cite{he2023sven}, which was manually labeled, covers only 9 Common Weakness Enumerations (CWEs) and comprises only 1.6k samples. This dichotomy presents a challenge: {how to acquire high-quality labeled \vd data at scale for training code LMs to detect security vulnerabilities}.

\paragraph*{Data duplication}
Furthermore, data duplication is prevalent in these datasets. Our analysis identified significant levels of exact copies and cloned vulnerabilities within the datasets. This is particularly problematic when one copy appears in the training set and another copy in the testing set, as performance metrics become unrepresentative of real-world performance and misrepresent the model's ability to generalize to unseen data. We found that up to 18.9\% test samples are leaked from the train set in some benchmarks. %

\smallskip
\noindent
\textbf{Limitation in existing evaluation metrics.}
Besides the data quality issue, the evaluation metrics used by current benchmarks fail to capture the practical utility of models at \vd.
\paragraph*{Accuracy} Many benchmarks report accuracy scores, but accuracy is not an appropriate metric for vulnerability detection, because of the base rate problem (vulnerabilities are rare in practice; most code is not vulnerable) and because of mismatches in class balance (the proportion of vulnerable samples in most research datasets does not match the ratio of vulnerable code in real life).
For instance, because most samples in realistic benchmarks are not vulnerable, it is possible to achieve high accuracy by always predicting ``not vulnerable''.
A high accuracy score does not necessarily signify effective detection of security vulnerabilities; it may simply reflect the accurate identification of non-vulnerable cases, or a bias towards predicting ``not vulnerable''. 
\paragraph*{F1} While the F1 score is widely perceived to be a better metric for assessing classification performance on imbalanced datasets, we argue that it is not appropriate in reality either.
The F1 score (the harmonic mean of precision and recall) reflects both false positives and false negatives by combining them into a single penalty.
Yet, for \vd tools in practice, the overwhelming majority of code is not vulnerable, so a critical challenge is preventing excessive false alarms.
The F1 score fails to reflect this asymmetry, so tools with a high F1 score may be useless in practice.

\smallskip
\noindent
\textbf{Proposed solution.} To address the above limitations, we propose a new \vd dataset and novel evaluation guidelines.

\paragraph*{New dataset} We introduce \dataset to tackle the limitations of existing datasets through rigorous data collection, data normalization, and data filtering process. 
We further introduce two rigorous labeling techniques: (i) \dataset-\textsc{NvdCheck} uses expert analysis from CVE entries, and 
(ii) \dataset-\textsc{OneFunc} utilizes unique changes within commits, ensuring high label accuracy. 
Consequently, \dataset not only ensures better label accuracy but also significantly reduces the possibility of data duplication, thereby offering a more realistic and less noisy \vd benchmark.
To this end, \dataset contains 6,968 vulnerable and 228,800 benign functions, covering 140 CWEs while maintaining similar accuracy as SVEN, marking a substantial advancement in both scale and accuracy compared to previous datasets.

\paragraph*{Novel evaluation guidelines} 
We propose novel evaluation guidelines, to ensure evaluation results will be predictive of the real-world performance of these tools. First, we suggest splitting samples chronologically to reflect the evolution of both vulnerabilities and coding patterns.  Chronological splitting also reduces the risks of data leakage. 
Second, we introduce the Vulnerability Detection Score (VD-S), a novel metric designed to measure how well vulnerability detectors will perform in practice.
VD-S measures the false negative rate, after the detector has been tuned to ensure the false positive rate is below a fixed threshold (e.g., 0.5\%).
Finally, we introduce a pair-wise evaluation method to assess the model's ability to distinguish between a vulnerable code sample and its benign (fixed) counterpart, 
offering deeper insights into models' vulnerability understanding.

\smallskip\noindent\textbf{Effectiveness of Code LMs in realistic vulnerability detection.} 
We evaluate seven code LMs with varied sizes, including the state-of-the-art open-source models like StarCoder2~\cite{lozhkov2024starcoder2} and the proprietary models from OpenAI~\cite{openai2024gpt4}, on \dataset.
Our findings paint a sobering picture of the current state of code LMs in vulnerability detection. Across various models and experimental setups, code LMs consistently performed poorly, as measured on the \dataset benchmark. 
This is in sharp contrast to prior evaluations on prior benchmarks, which reported seemingly good results.
For example, StarCoder2, which previously showed promising performance with a 68.26\% F1 score on BigVul, drastically underperformed in our assessment, achieving only a 3.09\% F1 score on \dataset. 

Our findings highlight the ineffectiveness of existing models in vulnerability detection and the misleading nature of previous benchmarks. 
This underscores the significance of \dataset in offering a more challenging and realistic evaluation environment, exposing the limitations of current models when confronted with real-world vulnerabilities. Additionally, the introduction of the Vulnerability Detection Score (VD-S) and pairwise evaluations further elucidated the challenges faced by the models.

Our attempts at enhancing model performance through advanced training techniques, such as class weights and contrastive learning, resulted in marginal improvements at best. We also evaluated state-of-the-art LLMs, including GPT-3.5 and GPT-4, where we employed zero-shot prompting and fine-tuning, hoping to leverage their massive pre-training and model capacity for enhanced performance. However, even these models cannot distinguish vulnerabilities from their similar yet benign counterparts.  In our pair-wise evaluation, GPT-4 with the chain-of-thought reasoning could not even outperform the random guessing. This indicates that we are a long way from being able to usefully detect security vulnerabilities with code language models, and fundamentally new approaches may be needed.
These findings also prompt a reassessment of the benchmarks used to gauge progress in the field, emphasizing the role of \dataset in advancing toward more realistic and rigorous evaluations.

To summarize, our contributions in this paper are as below:

\begin{itemize}
    \item We conducted an in-depth analysis of existing datasets, uncovering significant flaws, including poor data quality, low label accuracy, and a high incidence of data duplication.
    \item We developed \dataset, a new vulnerability dataset with high-quality, accurately labeled data and stringent de-duplication, to offer realistic training and testing data for code LMs.
    \item We introduced new evaluation guidelines, including the Vulnerability Detection Score (VD-S) and a pair-wise evaluation method, advancing the rigor of model assessment for vulnerability detection.
    \item We evaluated a range of code LMs with \dataset. Our evaluation demonstrates that, despite various attempts at optimization, the performance of current models significantly falls short of the requirements for real-world deployment, highlighting a pressing need for innovative approaches in the training of code LMs for vulnerability detection. We release our artifact at: \url{https://github.com/DLVulDet/PrimeVul}
\end{itemize}

\section{Background \& Challenges}

In this section, we will revisit the existing code-LM-based vulnerability detection models and study the core problems that prevent them from being promising and reliable for realistic deployment.

\subsection{Existing Data Collection Methods} 

\label{subsec: bg_data_label}
A vast amount of high-quality data is the key to successfully train deep neural networks. Earlier studies~\cite{li2018vuldeepecker, li2021sysevr} train deep-learning models using synthetic datasets such as SATE IV Juliet~\cite{okun2013report} and SARD~\cite{sard}. However, synthetic datasets do not adequately capture the complex and nuanced nature of vulnerabilities in real-world code~\cite{chakraborty2021deep}.
To address this, a series of benchmarks collect vulnerabilities from real-world, open-source software repositories~\cite{fan2020bigvul, chen2023diversevul, chakraborty2021deep, lu2021codexglue, zhou2019devign}. Existing vulnerability datasets suffer from one of the two following issues: (1) automated labeling is cheap but too coarse-grained, and (2) manual labeling is reliable but labor-intensive. We elaborate on this trade-off below.

\smallskip\noindent\textbf{Automated Labeling.}~The first strategy uses heuristics to automatically label all the samples so that the size of the dataset can be large enough for training deep neural networks. For example, BigVul~\cite{fan2020bigvul}, ReVeal~\cite{chakraborty2021deep}, CrossVul~\cite{nikitopoulos2021crossvul}, CVEfixes~\cite{bhandari2021cvefixes}, and DiverseVul~\cite{chen2023diversevul} follow this strategy. First, they collect vulnerability-fixing commits from open-source resources, such as the NVD database, Bugzilla, and Debian security trackers. Then, they label the before-commit version of changed functions as vulnerable, and their after-commit version and unchanged functions as nonvulnerable. This strategy assumes that the vulnerability-fixing commits only change vulnerable functions to fix the security flaw. In our label noise analysis (Section~\ref{subsec: problem_data_labeling}), we find that this is often not the case, which introduces wrong labels for vulnerable functions.

\smallskip\noindent\textbf{Manual Labeling.}~To improve the quality of the data labels, the second strategy is to involve human experts to verify the commits or functions manually. For example, the authors of SVEN~\cite{he2023sven} manually labeled 1,606 functions (half of them being vulnerable), from noisy datasets like BigVul and CrossVul. Unfortunately, even for the most experienced security experts, manually verifying the vulnerability labels of code samples is challenging and time-consuming. Therefore, the manually verified datasets, though having higher label accuracy, are not ideal for training the deep neural networks due to the limited diversity (e.g., SVEN only covers 9 CWEs) and a limited number of samples (e.g., SVEN only has 1.6k samples).

\subsection{Label Accuracy}
\label{subsec: problem_data_labeling}

In this section, we quantify the label accuracy of the existing benchmarks.

\subsubsection{Experiments} We analyze the label accuracy by randomly sampling 50 vulnerable functions from each benchmark and manually analyzing whether the function indeed contains security vulnerabilities. The manual analysis was performed by three of the authors, including two researchers with several years of experience in computer security and one senior security expert.

\smallskip\noindent\textbf{Human Judgement.}~
We follow the same method from Chen et al.~\cite{chen2023diversevul} to analyze whether each function is vulnerable and categorize nonvulnerable functions accordingly. Our human annotators comprehensively check the commit message that changed the sampled function, the function before and after the commit, the affiliated CVE, the NVD description, as well as the discussions among the developers in security issue trackers if available. Using these information, our human annotators confirm 1) whether the commit is related to fixing a security vulnerability, and 2) whether the sampled function is indeed vulnerable.

For nonvulnerable functions, we categorize them into 1) vulnerability spread across multiple functions, 2) other changes to make the fix consistent (relevant, but not vulnerable), and 3) irrelevant changes. In the first category, common scenarios where we cannot tell whether a single function is vulnerable include examples such as race condition, denial of service, or command injection that exploits multiple functions, etc. In the second category, we often observe changes to nonvulnerable functions as a side-effect of fixing vulnerable functions, e.g., reporting more errors outside the patched function, or changing the ways of calling the patched function. In the last category, the most common cases are changing spacing and functionality while fixing a different vulnerable function.

\smallskip\noindent\textbf{Majority Vote.}~As an improvement over Chen et al.~\cite{chen2023diversevul}, we use three human annotators with majority vote labeling for all datasets except SVEN. We label a function as vulnerable if at least two out of the three human annotators agree that the function is vulnerable. If any one of the three labelers is unsure about a function, the security expert will lead a discussion among annotators to achieve an agreement. Then, the final decision will be made through a majority vote.

\begin{table}[h]
\caption{
The label accuracy across existing vulnerability benchmarks and their comparison with two \dataset automated labeling techniques. Results show that our new labeling techniques achieve a high accuracy on par with SVEN which requires manual labeling. Moreover, \dataset contains 16.7 $\times$ as many vulnerable C/C++ functions as in SVEN.}
\centering
\resizebox{0.8\linewidth}{!}{
\begin{tabular}{l | c | c }
\toprule
\textbf{Benchmark} & \textbf{Manual} & \textbf{Correct (\%)} \\
\midrule

\textbf{SVEN~\cite{he2023sven}} & \mycmarkreverse & 94.0 \\

\textbf{CodeXGLUE~\cite{lu2021codexglue, zhou2019devign}} & \mycmarkreverse & 24.0 \\

\textbf{VulnPatchPairs~\cite{risse2023limits}} & \mycmarkreverse$^\dagger$ & 36.0 \\ \midrule

\textbf{BigVul~\cite{fan2020bigvul}} & \myxmarkreverse & 25.0$^*$ \\

\textbf{CrossVul~\cite{nikitopoulos2021crossvul}} & \myxmarkreverse & 47.8$^*$ \\

\textbf{CVEFixes~\cite{bhandari2021cvefixes}} & \myxmarkreverse & 51.7$^*$ \\

\textbf{DiverseVul~\cite{chen2023diversevul}} & \myxmarkreverse & 60.0$^*$ \\\midrule

\textbf{\dataset-\textsc{OneFunc}} & \myxmarkreverse & \textbf{86.0} \\
\textbf{\dataset-\textsc{NVDCheck}} & \myxmarkreverse & \textbf{92.0} \\

\bottomrule
\end{tabular}}
\begin{flushleft}
\small
$\dagger$ The VulnPatchPairs dataset takes pairs of functions from CodeXGLUE. The dataset does not involve further manual verification beyond its data resource, CodeXGLUE.\\
$^*$ Refers to label accuracy numbers in Chen et al.~\cite{chen2023diversevul}. %
\end{flushleft}
\vspace{-3mm}
\label{tab:label_accuracy}
\end{table}

\subsubsection{Results}
Table~\ref{tab:label_accuracy} summarizes our analysis results. We conduct accuracy analysis over three benchmarks that were originally constructed with some kind of manual labeling from prior work: CodeXGLUE, SVEN, and VulnPatchPairs. For benchmarks that use automated labeling, specifically, BigVul, CrossVul, CVEFixes, and DiverseVul, we refer to label accuracy numbers in Chen et al.~\cite{chen2023diversevul}. 
We will publicly release our label accuracy analysis results as part of our artifacts.

As shown in Table~\ref{tab:label_accuracy}, the benchmarks without manual verification have very low accuracy between 25\% and 60\%, for vulnerable functions. 
On the other hand, only one out of the three prior datasets that used manual labeling method has a high accuracy: SVEN has a 94\% label accuracy for vulnerable functions. The other two datasets, CodeXGLUE and VulnPatchPairs have low accuracy.

Surprisingly, the widely used dataset CodeXGLUE~\cite{lu2021codexglue} (a.k.a. Devign~\cite{zhou2019devign} dataset)
has a label accuracy of only 24\% for vulnerable functions, even though Zhou et al.~\cite{zhou2019devign} had recruited human annotators to label security-related commits. We find that their human annotation on security-related commits is highly inaccurate, such that many commits in CodeXGLUE do not fix security vulnerabilities. Moreover, they adopt the same automated labeling policy as BigVul~\cite{fan2020bigvul} to label all changed functions as vulnerable. The VulnPatchPairs~\cite{risse2023limits} dataset is derived from CodeXGLUE, and as a result, it is also inaccurate, with a label accuracy of only 36\%. Our labeling policy is more stringent than VulnPatchPairs. Similar to CodeXGLUE, we find that 24\% of functions from VulnPatchPairs are changed by some security-relevant commits, but the functions are not vulnerable. The vulnerability is either spread across multiple functions or the function is changed to make the fix consistent.

The most accurate dataset from prior work is SVEN, which has only 803 vulnerable functions (total $\sim$1.6k of vulnerable and non-vulnerable functions) across nine CWEs. In comparison, the other noisy datasets from Table~\ref{tab:label_accuracy} are much larger. Previous works have used different noisy datasets to fine-tune code LMs. Code LMs trained with these noisy datasets cannot be trusted for realistic deployment. When nearly half of the vulnerable samples in the training data have a wrong label, the quality of the trained model is rather concerning~\cite{Northcutt2021confident}.

Due to the space limit, we provide a more detailed label error analysis in the Supplementary Material I.A.

\subsection{Data Leakage}
\label{subsec: problem_data_duplication}

Data leakage has been identified as a significant issue in the area of machine learning for code. We consider two types of leakage: code copy and time travel.
Realistic evaluation of vulnerability detection models requires no data leakage to ensure their performances are reasonably measured. 
We study the data leakage issue of four most frequently used vulnerability benchmarks: BigVul, CVEFixes, CodeXGLUE, and DiverseVul, and they have been used by more than ten code-LM-based vulnerability detection models as training and evaluation material~\cite{feng2020codebert,guo2020graphcodebert,guo_unixcoder_2022,wang2021codet5,wang-etal-2023-codet5p,chen2023diversevul,fu2022linevul,steenhoek2023empirical,zhou2019devign, steenhoek2023language}. 

\smallskip\noindent\textbf{Data Splits.}~To study the data leakage issue specifically in the setting of fine-tuning for vulnerability detection, we need to create the train/validation/test splits. For CodeXGLUE, we directly take the original split from the benchmark~\cite{codexglue2019defect}. For BigVul, we take the public split~\cite{benjis2023bigvul} used by \cite{fu2022linevul, steenhoek2023empirical, steenhoek2023language}. For CVEFixes and DiverseVul, we follow the methodology introduced in \cite{chen2023diversevul} to randomly split the data with 80\%/10\%/10\%. 

\subsubsection{Code Copy}
One main reason for leakage is data duplication~\cite{allamanis2019duplication} since code data is highly repetitive~\cite{allamanis2014learning,allamanis2018survey},
and LMs are known to be good at memorizing the code text~\cite{rabin2023memorization}. Specifically, leaking exact copies across the training and evaluation set will inevitably inflate the evaluation performance.

\smallskip\noindent\textbf{Experiments.}~We study the exact copy of vulnerable functions. To identify the code copy, we exhaustively compare the vulnerable functions in the test set to all training samples. We normalize the formatting characters in the code samples to eliminate its noisy effect and then identify the exact copy if two strings are identical after the normalization.

\begin{table}[h]
\caption{
The statistics of data duplication in existing vulnerability detection benchmarks.}
\centering
\resizebox{0.7\linewidth}{!}{
\begin{tabular}{l | c | c }
\toprule
\textbf{Benchmark} & \textbf{De-dup} & \textbf{Copy (\%)} \\
\midrule

\textbf{BigVul~\cite{fan2020bigvul}} & \myxmark & 12.7  \\

\textbf{CVEFixes~\cite{bhandari2021cvefixes}} & \myxmark & 18.9  \\

\textbf{CodeXGLUE~\cite{lu2021codexglue}} & \myxmark & 0.6  \\

\textbf{DiverseVul~\cite{chen2023diversevul}} & \mycmark & 3.3  \\\midrule

\textbf{\dataset (Ours)} & \mycmark & \textbf{0.0}  \\

\bottomrule
\end{tabular}}

\label{tab:data_duplication}
\end{table}

\noindent\textbf{Results.}~Table~\ref{tab:data_duplication} reveals that existing benchmarks suffer from significant exact copy, up to 18.9\% of samples being duplicated. Unfortunately, this simple step is overlooked by prior work. Interestingly, we notice that, with hash-based deduplication, DiverseVul still has 3.3\% copies. This is mainly because they did not normalize formatting characters, and the same code with varied spacing will be mapped to distinct MD5 hashes, failing to be identified as copies.  Further, we manually check the root cause of such duplication, but we realize the reasons are varied. For example, in CodeXGLUE, we notice two different fixing commits targeting the same CVE~\footnote{\url{https://github.com/qemu/qemu/commit/71d0770}}~\footnote{\label{footnote:qemu_commit}\url{https://github.com/qemu/qemu/commit/902b27d}}. Consequently, this vulnerability gets sampled twice while one is in the training set and the other is in the test set. Differently, in BigVul, we identified exactly the same functions that being sampled twice before and after commits~\footnote{\url{https://github.com/php/php-src/commit/3798eb6}}~\footnote{\url{https://github.com/torvalds/linux/commit/6062a8d}}, having contradicting labels. Such noises will confuse the model during training and hurt the model's performance as a result.

\subsubsection{Time Travel}

Existing datasets also have the issue of time travel since they randomly separate functions into train, validation, and test sets. Consequently, it is possible to train on future data and test on past data. It is also possible to have the fixed nonvulnerable function in the training set, and the older vulnerable function in the test set.

In addition, after manual analysis, we find that many code samples from the same commit (example\footref{footnote:qemu_commit}) were randomly separated into train, validation, and test sets. This commit fixes multiple occurrences of the same CVE in different functions, in the same way. In general, developers tend to fix similar issues altogether before merging the changes into the main branch. However, training and testing with samples from the same commit is unrealistic and leaks information from the test time to the training time.
In a realistic setting,
the models are trained on the historical data to predict future samples.

\subsection{Evaluation in Practical Settings}
\label{subsec:limit_evaluation}

Nearly all code-LM-based vulnerability detection models use accuracy and F1 score for evaluation. Accuracy measures the proportion of correctly predicted samples (both vulnerable and benign) out of all samples, capturing overall model performance. The F1 score, on the other hand, is a harmonic mean of precision and recall, offering a balanced measure of the model's ability to identify vulnerable samples without excessively misclassifying benign samples. We argue that accuracy and F1 
fail to capture properties that developers care about in practice.
Rather, developers are generally concerned about (1) the detection error tradeoffs and (2) discriminative power over textually similar code samples across vulnerable and fixed patches.

\subsubsection{Detection Error Tradeoffs} Balance between false positives and false negatives is critical when deploying vulnerability detection tools. Developers rely on these tools not only to catch as many real vulnerabilities as possible (minimizing false negatives) but also to do so without overwhelming them with false alarms (minimizing false positives). This dual expectation reflects the practical tradeoffs in software development: missed vulnerabilities can lead to security breaches, while too many false alarms can lead to alert fatigue, potentially causing real issues to be ignored. F1 and accuracy offer little insight into the tradeoff between false negatives and positives.

\subsubsection{Textually Similar Pairs of Vulnerable \& Corresponding Patch} 
\label{subsubsec: problem_pairs}
Security fixes could involve only minor modifications to the code in many cases, such as adjusting buffer sizes, fixing data types, or adding security checks. These changes often result in a patched version of a function that is textually similar to its vulnerable counterpart. Code LMs, which primarily analyze the textual representation of code, struggle to differentiate between such closely related versions~\cite{ding2023concord,gu2024counterfeit,risse2023limits}. 
Only by evaluating models on these paired data, we can expose this weakness of code LMs.

\section{Our Benchmark: \dataset}
\label{sec:our_benchmark}

As demonstrated in Section~\ref{subsec: problem_data_labeling}, existing vulnerability benchmarks suffer from two major data quality issues: data duplication and low label accuracy. %
To address these shortcomings, we introduce a new vulnerability benchmark, \dataset. We propose a new automated data collection pipeline to produce high-quality samples with accurate labels.

\subsection{Data Merging and Thorough Data De-duplication}
\label{subsec: approach_data_dudup}
To build a large database, we start by merging all security-related commits and functions changed by them from BigVul~\cite{fan2020bigvul}, CrossVul~\cite{nikitopoulos2021crossvul}, CVEfixes~\cite{bhandari2021cvefixes}, and DiverseVul~\cite{chen2023diversevul}. We exclude data from Devign/CodeXGLUE~\cite{zhou2019devign, lu2021codexglue} because we discover that a large portion of its commits is unrelated to security issues during our annotation process.

We de-deduplicate code copies as well as functions with only formatting differences.
For each commit, we first normalize the changed functions before and after commits by stripping away characters such as spaces, tabs (``\textbackslash t''), newline characters (``\textbackslash n''), and carriage return characters (``\textbackslash r''). Then we compute the MD5 hash of both the pre- and post-commit versions of the changed function. If the pre- and post-commit versions of a function result in identical hash values, we regard this function as unchanged and discard it. Finally, we combine all remaining functions, and further de-duplicate the whole set by the MD5 hashes of the normalized functions. During the de-duplication, we maintain a unique set of hashes. If the hash of a normalized function is already in the set, we exclude it from further processing.

As a result, \dataset's thorough data de-duplication ensures that no vulnerable function from the training set can be leaked to the test set (Table~\ref{tab:data_duplication}).

\subsection{More Accurate Data Labeling}
\label{subsec: accuracte_data_labeling}

We propose two new labeling techniques: \dataset-\textsc{OneFunc} and \dataset-\textsc{NVDCheck}.

\smallskip\noindent\textbf{\dataset-\textsc{OneFunc}:} We notice that the previous labeling method has errors particularly when dealing with commits that modify multiple functions. Therefore, \dataset-\textsc{OneFunc} regards a function as vulnerable if it's the \emph{only} function changed by a security-related commit.

\smallskip\noindent\textbf{\dataset-\textsc{NVDCheck}:}~Since human experts have analyzed the CVEs in the NVD database, the vulnerability description in each CVE entry is a reliable reference to label vulnerable functions. We develop \dataset-\textsc{NVDCheck} as the following. First, we link security-related commits to their CVE numbers and the vulnerability description in the NVD database. 
We label a function as vulnerable if it satisfies one of the two following criteria: (1) NVD description explicitly mentions its name, or (2) NVD description mentions its \emph{file} name, and it is the only function changed by the security-related commit in that file.

After applying our two labeling techniques, we obtain two sets of vulnerable functions. Next, we merge the sets and de-duplicate
the functions again. We normalize the formatting characters in the functions, and compute their MD5 hashes to identify and remove duplicates. Subsequently, we label the post-commit versions of these identified vulnerable functions, as well as all other unchanged functions within the same commits, as benign. Only a subset of commits from the merged database mentioned in Section~\ref{subsec: approach_data_dudup}, meet the criteria for labeling by our techniques. Commits without any function that matches these criteria are excluded from our dataset.

Our pipeline results in a collection of 6,968 vulnerable and 228,800 benign functions across 755 projects and 6,827 commits. To assess our labeling accuracy, we conducted a manual review following the same process used in Section~\ref{subsec: problem_data_labeling}, with our results presented in Table~\ref{tab:label_accuracy}. The most accurate prior dataset SVEN has only 417 vulnerable functions in C/C++, and 386 vulnerable functions in Python.
Our \dataset{} dataset not only matches the label accuracy of SVEN but also significantly expands the collection of vulnerable C/C++ functions by 16.7$\times$ compared to SVEN. \dataset{} is diverse, containing 140 CWEs (15.6$\times$ of SVEN).

\vspace{-2mm}
\begin{table}[h]
\caption{Statistics of \dataset}
\centering
\resizebox{0.9\linewidth}{!}{
\begin{tabular}{c|c|c|c|c}
\toprule
\multirow{2}{*}{\textbf{Split}} & \multicolumn{2}{|c|}{\textbf{All}} & \multicolumn{2}{|c}{\textbf{Paired}}\\
\cmidrule{2-5}
 & \textbf{\# Vuln.} & \textbf{\# Benign}  & \textbf{\# Vuln.} & \textbf{\# Benign}  \\\midrule
 
\textbf{Train} & 5,574 & 178,853 & 4,354 & 4,354 \\
\textbf{Dev} & 699 & 24,731 & 562 & 562 \\
\textbf{Test} & 695 & 25,216 & 564 & 564 \\
\bottomrule
\end{tabular}}
\vspace{-3mm}
\label{table:data_stats}
\end{table}
\section{New Evaluation Guidelines}
\label{sec:evaluation_guidelines}

As discussed in Section~\ref{subsec: problem_data_duplication}, we need new methods to properly evaluate vulnerability detection models in deployment settings. This section proposes new evaluation guidelines.

\subsection{Temporal Splits}
\label{subsec:time-aware_splits}

To minimize the data leakage issue and formulate a realistic train-evaluate setup for vulnerability detection, we split the train/validation/test set of \dataset according to the commit date of the samples. Concretely, we find the original commit for each sample and collect the time of that commit, tying it with the sample. Then, we sort the samples according to the commit, where the oldest 80\% will be the train set, 10\% in the middle will be the validation set, and the most recent 10\% will be the test set. We also make sure that the samples from the same commit will not be split into different sets. This ensures that the vulnerability detection model is trained using past data and tested over future data.

\subsection{More Realistic and Challenging Evaluation}
\label{subsec: realistic_evaluation}

\subsubsection{Vulnerability Detection Score} 
\label{subsubsec:vds}
The primary goal in vulnerability detection is to catch as many real vulnerabilities as possible (can be measured by False Negative Rate, or $FNR$, where we expect low $FNR$).
Meanwhile, from a practical perspective, a certain level of false positives can be manageable within the development workflow, without causing alert fatigue (typically captured by False Positive Rate, or $FPR$, where a lower rate is better). 
Therefore, a metric that focuses on minimizing the false negative rate within a tolerable level of false positives is essential.

To this end, we propose Vulnerability Detection Score (VD-S), that evaluates the False Negative Rate of a vulnerability detector within an acceptable False Positive Rate, %
i.e., $FNR$ @ $(FPR \leq r)$, where $r \in [0\%, 100\%]$ is a configurable parameter.
In this paper, we choose a tolerance rate $r = 0.5\%$ to perform the evaluation in Section~\ref{sec: results}.

\subsubsection{Paired Functions and Pair-wise Evaluation}
\label{subsubsec:paired_function_evaluation}

As discussed in Section~\ref{subsubsec: problem_pairs}, evaluating the models on paired functions---vulnerable and benign versions of code---could potentially reveal whether a model merely relies on superficial text patterns to make predictions without grasping the underlying security implications, indicating areas where the model needs improvement to reduce the false positives and false negatives. 

We collected 5,480 such pairs in \dataset, significantly larger than existing paired datasets~\cite{risse2023limits, he2023sven}. Concretely, we match the vulnerable functions with their patches in \dataset to construct such pairs. As we show in Table~\ref{table:data_stats}, the paired vulnerable functions are fewer than all vulnerable functions, since not all vulnerable functions have a patch (e.g., a patch could delete the vulnerable function), and we only include those challenging pairs sharing at least 80\% of the string between the vulnerable and benign version.

Accordingly, we also propose a pair-wise evaluation method. The core idea is to evaluate the model's predictions on the entire pair as a single entity, emphasizing the importance of correctly identifying both the presence and absence of vulnerabilities in a textually similar context, while recording the model's concrete predicting behaviors.

We define four outcomes of the pair-wise prediction:
\begin{itemize}[leftmargin=*]
    \item \emph{Pair-wise Correct Prediction (P-C)}: The model \emph{correctly} predicts the ground-truth labels for both elements of a pair.
    \item \emph{Pair-wise Vulnerable Prediction (P-V)}: The model \emph{incorrectly} predicts both elements of the pair as \emph{vulnerable}.
    \item \emph{Pair-wise Benign Prediction (P-B)}: The model \emph{incorrectly} predicts both elements of the pair as \emph{benign}.
    \item \emph{Pair-wise Reversed Prediction (P-R)}: The model \emph{incorrectly} and inversely predicts the labels for the pair.
\end{itemize}

\section{Experimental Results}
\label{sec: results}

Considering the better data labeling and closer alignment with the real-world data distribution in \dataset, coupled with the introduction of new evaluation techniques, we reassess code LM's performance using  \dataset to gauge their performance in a more realistic setting. To this end, we evaluate the following three Research Questions:
\begin{itemize}[leftmargin=*]
    \item \textbf{RQ1.} How do open-source code LMs perform on \dataset? (\Cref{subsec:RQ1})
    \item \textbf{RQ2.} Can employing more advanced training techniques enhance the performance of code LMs in detecting vulnerabilities? (\Cref{subsec:RQ2})
    \item \textbf{RQ3.} Can larger language models (LLMs) improve vulnerability detection performance? (\Cref{subsec:RQ3})
\end{itemize}

\subsection{Study Subject}

\smallskip\noindent\textbf{Datasets.} We mainly use \dataset to conduct our experiments. In RQ1, we additionally use BigVul~\cite{fan2020bigvul} as a case study of existing benchmarks due to its popularity~\cite{fu2022linevul, steenhoek2023empirical, steenhoek2023language, benjis2023bigvul, chen2023diversevul}. By comparing it to \dataset, we illustrate the impact of its limitation on training and evaluating code-LM-based \vd models. For RQ2 and RQ3, we will focus on \dataset to improve code LMs performances for \vd.

\begin{table}[h]
\caption{
The code LMs we will study in this section.}
\centering
\resizebox{0.9\linewidth}{!}{
\begin{tabular}{l | c | c | c}
\toprule
\textbf{Model} & \textbf{Parameters} & \textbf{Arch} &\textbf{Methods} \\\midrule
\multirow{1}{*}{{\textsc{CodeT5~\cite{wang2021codet5}}}} & 60 M & Enc-Dec & Fine-tune \\\midrule
\multirow{1}{*}{{\textsc{CodeBERT~\cite{feng2020codebert}}}} & 125 M & Encoder & Fine-tune \\\midrule
\multirow{1}{*}{{\textsc{UnixCoder~\cite{guo_unixcoder_2022}}}} & 125 M & Encoder & Fine-tune \\\midrule
\multirow{1}{*}{{\textsc{StarCoder2~\cite{lozhkov2024starcoder2}}}} & 7 B & Decoder & Fine-tune \\\midrule
\multirow{1}{*}{{\textsc{CodeGen2.5~\cite{nijkamp2023codegen2}}}} & 7 B & Decoder & Fine-tune \\\midrule
\multirow{3}{*}{{\textsc{GPT-3.5~\cite{ouyang2022training}}}} & \multirow{3}{*}{$>100$B} & \multirow{3}{*}{Decoder} & Fine-tune \\
&&&Few-shot Prompt\\
&&&Chain-of-thought\\
\midrule
\multirow{2}{*}{{\textsc{GPT-4~\cite{openai2024gpt4}}}} & \multirow{2}{*}{$>100$B} & \multirow{2}{*}{Decoder} & Few-shot Prompt\\
&&&Chain-of-thought\\
\midrule
\end{tabular}}
\vspace{-2mm}
\label{tab:studied_models}
\end{table}

\smallskip\noindent\textbf{Models.} We will study seven code LMs with varied sizes regarding their capabilities for \vd, as shown in Table~\ref{tab:studied_models}. Specifically, we will fine-tune all open-source models in RQ1 and study the advanced training techniques using UnixCoder in RQ2. In RQ3, we prompt GPT-3.5 and GPT-4 with two settings: (1) two-shot examples (2) and chain-of-thoughts~\cite{wei2022chain} reasoning, and we also fine-tune GPT-3.5 on a subset of \dataset using the OpenAI API.

\smallskip\noindent\textbf{Experimental Settings.} For all experiments with open-source models, we implement our fine-tuning framework following the existing benchmarks~\cite{lu2021codexglue, chen2023diversevul}, where we will use the learning rate of $2 \times 10^{-5}$ for all the fine-tuning. For smaller models with less than 7B parameters, we will fine-tune the models for 10 epochs, and for 7B models, we fine-tune for four epochs. For GPT-3.5, we just fine-tune for one epoch, due to the limited budget. We load the model weights from Hugging Face Models\footnote{\url{https://huggingface.co/models}}. All training tasks are conducted on a cluster with NVIDIA A100 GPUs (80 GB). Experiments with OpenAI models are performed through API using greedy decoding.

\subsection{RQ1: Performance of Open-Source Code LMs on \dataset}
\label{subsec: empr_setup}
\label{subsec:RQ1}

In this RQ, we fine-tune open-source code LMs and evaluate their performances on \dataset. 
To illustrate the %
limitations of existing \vd benchmarks on the model training and evaluation, we reproduce the code LMs' performances on BigVul~\cite{fan2020bigvul}, using which as a direct comparison to \dataset. Specifically, we additionally fine-tune code LMs on BigVul, and evaluate them on both BigVul and \dataset. 

\smallskip\noindent\textbf{Results.}~The empirical results are shown in Table~\ref{tab:main}. The rows of ``Train=PV" and ``Test=PV" in each section are results for code LMs fine-tuned and evaluated on \dataset, and rows of ``Train=BV" are the comparative results for code LMs fine-tuned with BigVul.

\begin{table}[ht]
\caption{
Performance of code-LM-based vulnerability detection models in different settings. 
}
\addtolength{\tabcolsep}{-0.4em}
\centering
\resizebox{\linewidth}{!}{
\begin{tabular}{l | c | c | r  r  r | r  r  r  r }
\toprule
\textbf{Model} & \textbf{Train} & \textbf{Test} & \textbf{Acc $\uparrow$} & \textbf{F1 $\uparrow$} & \textbf{VD-S $\downarrow$} & \textbf{P-C $\uparrow$} & \textbf{P-V $\downarrow$} & \textbf{P-B $\downarrow$} & \textbf{P-R $\downarrow$} \\
\midrule
\midrule
\multirow{3}{*}{{\textsc{CT5}}} & \multirow{2}{*}{BV} &\colorgray BV& \colorgray 95.67 &\colorgray 64.93 & \colorgray 77.30 &\colorgray 24.98 &\colorgray 50.90 &\colorgray 22.79 &\colorgray 1.33\\

&  &\colorblue PV&\colorblue 97.00 & \colorblue 5.82 &  \colorblue 95.97 &\colorblue  0.18 &\colorblue 3.01 &\colorblue 96.10 &\colorblue 0.71\\\cmidrule{2-10}

& PV  &\colorblue PV&\colorblue 96.67 & \colorblue 19.7 & \colorblue 89.93 &\colorblue 1.06 &\colorblue 12.94 &\colorblue 84.75 &\colorblue 1.24\\

\midrule
\midrule
\multirow{3}{*}{{\textsc{CB}}} & \multirow{2}{*}{BV} &\colorgray BV& \colorgray 95.57  & \colorgray 62.88 & \colorgray 81.77 & \colorgray 22.60 & \colorgray 48.34 & \colorgray 27.83 & \colorgray 1.23\\

& &\colorblue PV&\colorblue 97.04 & \colorblue 4.49 & \colorblue 95.54 &\colorblue  0.35 &\colorblue 1.95 &\colorblue 96.99 &\colorblue 0.71\\\cmidrule{2-10}

& PV  &\colorblue PV&\colorblue 96.87 & \colorblue 20.86 & \colorblue 88.78 &\colorblue 1.77 &\colorblue 11.35 &\colorblue 86.17 &\colorblue 0.71\\

\midrule

\multirow{3}{*}{{\textsc{UC}}} & \multirow{2}{*}{BV} &\colorgray BV& \colorgray 96.46 & \colorgray 65.46 & \colorgray 62.30 & \colorgray 39.60 & \colorgray 23.74 & \colorgray 33.24 & \colorgray 3.42\\
&  &\colorblue PV&\colorblue  \colorblue 97.27 & \colorblue 1.94 & \colorblue 95.11 &\colorblue  0.35 &\colorblue 0.35 &\colorblue 98.76 &\colorblue 0.53\\\cmidrule{2-10}
& PV  &\colorblue PV&\colorblue 96.86 & \colorblue 21.43 & \colorblue 89.21 &\colorblue 1.60 &\colorblue 12.06 &\colorblue 85.11 &\colorblue 1.24\\\midrule\midrule

\multirow{3}{*}{{\textsc{SC2}}} & \multirow{2}{*}{BV} & \colorgray BV& \colorgray 96.20 & \colorgray 68.26 & \colorgray 69.14 & \colorgray 35.23 & \colorgray 41.98 & \colorgray 20.61 & \colorgray 2.18 \\
&  & \colorblue PV & \colorblue 97.09 & \colorblue 3.09 & \colorblue 96.83 & \colorblue 0.89 & \colorblue 0.89 & \colorblue 97.70 & \colorblue 0.53 \\\cmidrule{2-10}
& PV  & \colorblue PV & \colorblue 97.02 & \colorblue 18.05 & \colorblue 89.64 & \colorblue 2.30 & \colorblue 8.16 & \colorblue 88.30 & \colorblue 1.24 \\\midrule

\multirow{3}{*}{{\textsc{CG2.5}}} & \multirow{2}{*}{BV} & \colorgray BV& \colorgray 96.57  & \colorgray 67.30  & \colorgray61.73  & \colorgray 40.84  & \colorgray 26.02 & \colorgray 29.63 & \colorgray 3.51\\

&  & \colorblue PV & \colorblue 97.23  & \colorblue 1.91 & \colorblue 95.68 & \colorblue 1.24  & \colorblue 0.00 & \colorblue 98.76 & \colorblue 0.00 \\\cmidrule{2-10}

& PV  & \colorblue PV & \colorblue 96.65 & \colorblue 19.61 & \colorblue 91.51 & \colorblue 3.01 & \colorblue 10.82 & \colorblue 84.22 & \colorblue 1.95\\ %
\bottomrule 
\end{tabular}}

{
\smallskip
{
\begin{flushleft}
CT5: CodeT5, CB: CodeBERT, UC:UnixCoder, SC2: StarCoder2, CG2.5: CodeGen2.5. BV: BigVul, PV: \dataset. VD-S is the Vulnerability Detection Score defined as false negative rate (FNR) @ false positive rate (FPR) $\leq$ 0.5\% in Section~\ref{subsubsec:vds}. P-C, P-V, P-B, and P-R are the metrics defined in Section~\ref{subsubsec:paired_function_evaluation} to evaluate the models on paired functions.%
\end{flushleft}
} }
\vspace{-3mm}
\label{tab:main}
\end{table}
\smallskip

\subsubsection*{Finding-RQ1.1: Code LMs' performance is overestimated on a prior benchmark and perform poorly on \dataset}
The comparative performance evaluation of code LMs between \dataset and BigVul lays bare a startling truth: the proficiency of these models is greatly overestimated by benchmarks like BigVul, which fail to mimic the complexity of real-world vulnerabilities. 
For example, while StarCoder2 shows a commendable F1 score of 68.26\% on BigVul, it plummets to a paltry 3.09\% on \dataset. This precipitous drop is not an isolated case but a trend observed across all models, exemplified by the observed false negative rates. 

In addition, even when code LMs are fine-tuned on \dataset, they fail to achieve the same level of performance as on BigVul. For instance, when trained on \dataset, StarCoder2’s performance shows only a modest improvement in F1 score from 3.09\% to 18.05\%, which is still markedly lower than the 68.26\% F1 score achieved on BigVul.
This persistent underperformance, even after fine-tuning, suggests that the models cannot effectively learn from the more complex and realistic distribution of vulnerabilities in \dataset. This stark discrepancy confirms that code LMs trained on existing benchmarks may develop a false sense of security due to the benchmarks' failure to capture the intricate and diverse nature of vulnerabilities found in the wild.

\smallskip\subsubsection*{Finding-RQ1.2: Vulnerability Detection Score (VD-S) offers a more concrete sense of realistic performance} The VD-S emerges as a pivotal metric, capturing the essence of a model's capability in real-world settings, where balancing the FNR and FPR is crucial. For example, CodeBERT reports a high accuracy of 96.86\% and a satisfactory F1 of 62.88\% on BigVul. Relying on these metrics, CodeBERT seems to be an acceptable candidate for detecting vulnerabilities. However, its astonishing false negative rate of 81.77\% (as reported by VD-S) far exceeds the realistic tolerable limits, revealing that it is actually useless in practice, overturning the conclusions drawn from the accuracy and F1. Code LMs' VD-S on \dataset is even more concerning than on BigVul, highlighting the models' limitations in accurately identifying true vulnerabilities, a critical factor for real-world applications where missing a single vulnerability can have serious repercussions.
 
In addition, we realize that VD-S is not necessarily correlated with accuracy or F1. The detachment of VD-S from traditional metrics also signals a shift in how we should evaluate code LMs for \vd.

\smallskip\subsubsection*{Finding-RQ1.3: Code LMs are weak at differentiating vulnerabilities from their similar but benign counterparts} 
As we have discussed in Section~\ref{subsubsec: problem_pairs} and \ref{subsubsec:paired_function_evaluation}, pair-wise evaluation not only offers a lens into the precision of code LMs but also serves as a stress test for their real-world application viability. These metrics crucially reveal whether a model has truly learned to identify security vulnerabilities or is merely recognizing patterns without comprehending the implications—a distinction that's vital for the deployment of code LMs as a reliable security tool. 

However, our pair-wise evaluation uncovers a significant deficiency in this aspect. Our results reveal that code LMs frequently misclassify both functions in a pair as vulnerable (P-V) or benign (P-B), indicating an overreliance on textual patterns rather than a substantive understanding of the code's security context. For instance, StarCoder2 reports only 2.30\% cases to correctly label both elements in paired functions while misclassifying 4$\times$ more cases as both vulnerable and 88.30\% pairs as both benign, demonstrating their difficulty in recognizing the vulnerable patterns and subtle distinctions that a patch can introduce. Such unreasonable behaviors undermine the trustworthiness of these models in realistic deployment. This insight is also crucial as it emphasizes the need for models that go beyond surface-level text comprehension to grasp deeper semantic implications of code changes for reliable vulnerability detection.

\begin{finding}
Code LMs' significant underperformance on \dataset highlights their limitations when faced with realistic, diverse, and challenging vulnerabilities. 
\end{finding}

\subsection{RQ2: Exploring to Improve the Performance on \dataset}
\label{subsec:RQ2}
Given code LMs' poor performance on \dataset, we decided to delve deeper into the training process, trying to figure out whether more advanced training techniques could help code LMs achieve promising performance on \dataset.

To this end, we perform analysis to inspect both the challenging samples within \dataset and monitor the models' behaviors during the inference time, carefully studying the failing predictions from these models. After these analyses and a comprehensive literature review, we decided to explore two advanced training techniques that have shown effectiveness in helping binary classifications.

\smallskip\subsubsection{Exploration-1: Class Weights}

When we delve deeper into analyzing the experimental results, one of the notable differences between BigVul and \dataset is the ratio of vulnerable samples. As we have mentioned in Section~\ref{subsec: problem_data_labeling}, a lot of samples are mislabeled as vulnerable in BigVul and other resources that constitute \dataset. After applying our novel labelers, (Section~\ref{subsec: accuracte_data_labeling}), the ratio of vulnerabilities significantly decreases. Therefore, we suspect whether the significantly more imbalanced ratio hinders the learning process~\cite{johnson2019imbalance}. 

To verify our assumptions, we implement the weighted loss similar to Chen et al.~\cite{chen2023diversevul} and integrate it into our fine-tuning framework. The general idea is to give a higher penalty when models make mistakes on the rare class (i.e., the vulnerable samples) by up-weighting the loss value for this class in the cross entropy loss. With a higher weight on the rarer class, the imbalance ratio will be less harmful, since the model pays comparable attention to both classes when optimizing the loss. To find an optimal weight, we tried several different values: besides the standard binary classification with equal weights (weight of vulnerable class = 1), we further explored upweighting the loss for the vulnerable class for 5, 20, 30, and 40 times. Note that the vulnerable to benign ratio in \dataset is roughly 1:32.

\begin{table}[h]
\caption{
The impact of class weights during training.}
\centering
\resizebox{\linewidth}{!}{
\begin{tabular}{c | r  r  r | r  r  r  r }
\toprule
\textbf{Weight} & \textbf{Acc$\uparrow$} & \textbf{F1$\uparrow$} & \textbf{VD-S$\downarrow$} & \textbf{P-C$\uparrow$} & \textbf{P-V$\downarrow$} & \textbf{P-B$\downarrow$} & \textbf{P-R$\downarrow$} \\
\midrule
\midrule
\textbf{1} & 96.86 & 21.43 & 89.21  & 1.60 & 12.06 & 85.11 & 1.24\\\midrule

\textbf{5} & 96.24 & 25.29 & 90.65 & 1.77 & 18.97 & 78.55 & 0.71 \\

\textbf{20} & 95.28 &24.26 & 88.92 & 0.89 & 25.71 & 72.16 & 1.24\\

\textbf{30}& 96.14 & 24.49 & 90.07 & 2.13 & 18.09 & 78.01 & 1.77\\
\textbf{40}& 95.99 & 26.28 & 88.49 & 1.42 & 22.52 & 74.82 & 1.24\\

\bottomrule
\end{tabular}}

\label{tab:class_weights}
\end{table}

\subsubsection*{Findings-RQ2.1: Class weights do not fundamentally improve code LMs' performance on \dataset} Similar to Chen et al.~\cite{chen2023diversevul}, we observed an increased F1 when applying class weights, though not as significant as theirs due to the difficulty of \dataset. However, VD-S scores warn us that such a marginal improvement is far from promising in the realistic scenario: FNR is oscillating around 90\% across different weights, and such models could not be trusted to detect security flaws. These results, though disappointing, help us exclude the potential impact of the class imbalance, providing convincing evidence regarding the difficulty of \dataset and the struggle of code LMs in detecting realistic vulnerabilities.

\smallskip\subsubsection{Exploration-2: Contrastive Learning}

Contrastive learning has been proven effective at learning better-quality representations of text and code~\cite{gao2021simcse, guo_unixcoder_2022, ding2023concord} since they are able to decrease the cosine similarity among semantically different samples, and consequently help to improve the models' performance in downstream classification tasks. Therefore, we hope to study whether contrastive learning could help code LMs to achieve promising performance on \dataset.

Different from the existing works~\cite{guo_unixcoder_2022, ding2023concord} which apply contrastive learning at the pre-training phase, we enforce such ``contrasting'' signals together with the classification objective. Specifically, code LMs will be fine-tuned to both classify vulnerabilities and contrast representations with distinct semantics. We implement the objective according to Gao et al.~\cite{gao2021simcse} (referred to as CLR), where the model is trained to maximize the representation similarity between each sample and the perturbed version of itself and minimize the similarity between two randomly chosen samples. The perturbation is directly applied to the code representation through dropouts~\cite{srivastava2014dropout} in the Transformer model.

\begin{table}[h]
\caption{
The impact of contrastive learning during fine-tuning UnixCoder for vulnerability detection.}
\centering
\resizebox{\linewidth}{!}{
\begin{tabular}{l | r  r  r| r  r  r  r   }
\toprule
\textbf{Finetune} & \textbf{Acc$\uparrow$} & \textbf{F1$\uparrow$} & \textbf{VD-S$\downarrow$} & \textbf{P-C$\uparrow$} & \textbf{P-V$\downarrow$} & \textbf{P-B$\downarrow$} & \textbf{P-R$\downarrow$} \\
\midrule
\midrule
\textbf{CLS} & 96.86 & 21.43 & 89.21  & 1.60 & 12.06 & 85.11 & 1.24\\\midrule

\textbf{\quad + CLR} & 96.83 & 21.73 & 90.07  & 1.77 & 11.35 & 85.82 & 1.06 \\\midrule

\textbf{\quad + CA-CLR} & 96.64 & 24.46 & 90.50  & 2.30 & 15.96 & 81.38 & 0.35 \\

\bottomrule
\end{tabular}}
\label{tab:clr_study}
\end{table}

\subsubsection*{Findings-RQ2.2: Contrastive learning fails to significantly improve Code LMs' performance on \dataset} Unfortunately, as shown in Table~\ref{tab:clr_study}, we could not see a significant difference by adding the CLR objective. We further analyze the results to see what might go wrong. One notable misalignment we notice is that, since CLR from Gao et al.~\cite{gao2021simcse} is not crafted for classification tasks, it will distinguish any two samples regardless of whether their labels are the same. Therefore, we further improve CLR to be a second approach, called Class-aware Contrastive Learning (CA-CLR), which will only minimize the similarity between samples with different labels.

This time, we see a more notable improvement over F1 and P-C by applying CA-CLR (Table~\ref{tab:clr_study}). However, the performance change is still marginal. This result empirically demonstrates code LMs' inherent incapability to detect vulnerabilities. It is not only because the representations are too similar to draw the classification boundary, but these models fail to identify the vulnerable patterns, so that, even if enlarging the cosine distance among representations through contrastive learning, code LMs still fail to draw the classification boundary correctly.

\begin{finding}
Advanced training techniques for binary classification, such as contrastive learning and class weights, could not particularly improve code LMs' performance on \dataset, highlighting its realistic difficulty.
\end{finding}

\subsection{RQ3: Larger Code LMs on \dataset}
\label{subsec:RQ3}

After getting unsatisfying results with advanced training techniques, we started to question whether the models we tried so far have too few parameters to solve such a challenging benchmark. Therefore, we decided to explore the state-of-the-art large language models (LLM) to see whether significantly more parameters could bring a performance.

We perform experiments using OpenAI GPT models: GPT-3.5 and GPT-4. Considering the cost, we only evaluate these models on the paired functions of \dataset, since it has much fewer samples than the full set while representing the more challenging scenarios in reality. For GPT-3.5, we experiment with three settings: two-shot prompting, chain-of-thought prompting~\cite{wei2022chain}, and fine-tuning. We fine-tune GPT-3.5 for one epoch on all vulnerabilities plus three times more randomly sampled benign samples from the train split of \dataset, as fine-tuning on the full training set will run out of the budget. For GPT-4, we only consider two-shot and chain-of-thought prompting since its fine-tuning API has not been released yet. More details about the prompt template will be discussed in Supplementary Material I.B.

\begin{table}[h]
\caption{
Results of OpenAI GPT models on \dataset paired functions. }
\centering
\resizebox{\linewidth}{!}{
\begin{tabular}{c | c | r  r  r  r }
\toprule
\textbf{Model} & \textbf{Method} & \textbf{P-C $\uparrow$} & \textbf{P-V $\downarrow$} & \textbf{P-B $\downarrow$} & \textbf{P-R $\downarrow$} \\
\midrule
\midrule
\multirow{3}{*}{{\textsc{\textsc{GPT-3.5}}}} & Two-shot & 5.67 & 13.83 & 77.84 & 2.66\\\cmidrule{2-6}
& CoT & 6.21 & 4.79 & 83.51 & 5.50\\\cmidrule{2-6}
& Fine-tune & 1.24 & 5.32 & 90.96 & 2.48 \\
\midrule
\midrule
\multirow{3}{*}{{\textsc{\textsc{GPT-4}}}} & Two-shot & 5.14 & 71.63 & 21.45 & \textbf{1.77}\\\cmidrule{2-6}
& CoT & 12.94 & 54.26 & 24.47 & 8.33\\
\midrule
\midrule
\textsc{Random Guess}& - & 22.70 & 26.24 & 26.42 & 24.65\\
\midrule
\bottomrule
\end{tabular}}

\label{tab:gpt_result}
\end{table}

Results are shown in Table~\ref{tab:gpt_result}. In general, GPT-3.5 and GPT-4 outperform open-source models for the pair-wise evaluation even in the basic two-shot prompting setting, and chain-of-thought reasoning further pushes the performance boundary. However, we realize that such performance is actually no better than a random guess since the majority of the pairs in \dataset still cannot be distinguished by these large SOTA code LMs, which might indicate the fundamental weaknesses of code LMs to differentiate subtle vulnerabilities from their benign versions. Furthermore, when we fine-tune GPT-3.5, we notice that the model is strongly biased by the 1:3 vulnerable to benign ratio and reports even lower performance than the prompting approaches, showing a red flag that even such a large LM (LLM) still fails to capture the vulnerable patterns but take shortcuts from data instead.

\begin{finding}
Even the state-of-the-art OpenAI models could not achieve a reliable performance on \dataset, calling for fundamentally novel approaches to improve the task.
\end{finding}

\vspace{2mm}
\section{Discussions \& Threats to Validity}
\label{sec: discussion}

\noindent\textbf{Discussion.}~Our study of code LMs in the realm of vulnerability detection (\vd) reveals that they do not perform well enough for real-world applications. Prior evaluations looked promising, but our work reveals 
their subtle issues:
data quality problems, misleading metrics, and methodology that poorly matches the way in which these models would be used in practice. 
We highlight below several key areas where current code LMs fall short.

\paragraph{Need for More Context} Prior work formulates the problem as: given the code of a single function, determine whether that function contains a security vulnerability. However, this may be asking an impossible question. Determining whether code is vulnerable generally depends on information about other components of the system as well, such as whether inputs to the function have already been sanitized, how outputs will be used, or what invariants are established by the rest of the system. This focus on function-level analysis without the consideration of other contexts (such as interprocedural data flows) would make it difficult for even a human to detect vulnerabilities, let alone a model.  We recommend that the problem be reformulated so that the model also has access to a broader context. To enable such a process, \dataset maintains the metadata for the included commits, providing resources to extract relevant contexts.  

\paragraph{Augmenting Security Awareness} Our empirical results, particularly the shortcomings of code LMs in pair-wise evaluations, suggest that these models make decisions primarily based on textual similarity, without considering the underlying root causes or fixes of the vulnerabilities. 
We suggest researchers explore ways to teach code LMs about security concepts, such as pre-training methods inspired by how we teach human software developers about security, or ways to build hybrid systems that combine LMs with traditional security analysis or program analysis tools~\cite{khare2023understanding, li2024enhancing}. 

\paragraph{Teaching the model to reason about \vd}
Finally, posing vulnerability detection as a binary classification problem and teaching the Code LMs accordingly might be too simplistic. This approach banks on the slim hope that a lone summary token or a condensed representation can embody all the intricacies of code vulnerabilities—such expectation might be overly optimistic. Instead, we should decompose the \vd problem into digestible sub-problems and teach the model to reason about each step to reach a conclusion~\cite{sun2024llm4vuln}. The slight yet encouraging progress observed with the chain of thought experiments in~\Cref{tab:gpt_result} shows some promise in this direction.

\smallskip\noindent\textbf{Threats to validity.}~Even with our stringent labeling methods, label accuracy is less than 100\% (see Table~\ref{tab:label_accuracy}), so there is still a small portion of mislabeled data in \dataset. 
However, given the small percentage of mislabeling, we believe all the reported results will still hold good. 

For the proposed evaluation metric VD-S, there is a configurable parameter $r$ to control the maximum false positive rate.
The practically acceptable value $r$ might vary for different scenarios, changing the exact value of VD-S, but we expect our general conclusions to hold. 

For experiments with OpenAI models, we only reported results with default settings, which could vary slightly by changing the hyperparameters. %
However, we do not expect hyperparameter tweaking would change the conclusion.

\section{Related Work}

Throughout the paper, we have reviewed many related works. 
Here, we provide a summary of the remaining ones and compare them with our contribution.

\paragraph*{Code-LM-based Vulnerability Prediction} Two primary methods to use Code LMs for \vd are fine-tuning and prompting. 
Fine-tuning adds a randomly initialized binary classification head to the language model and jointly optimizes all weights based on ground-truth labels. 
Various approaches, such as encoder-decoder Transformers~\cite{ahmad-etal-2021-unified, wang2021codet5}, encoder-only Transformers~\cite{feng2020codebert, guo_unixcoder_2022}, and decoder-only Transformers~\cite{li2023starcoder, lozhkov2024starcoder2,nijkamp2023codegen2} have been used for fine-tuning. 
On the other hand, prompting-based methods~\cite{khare2023understanding,ullah2023can,sun2024llm4vuln,li2024enhancing} rely on LLMs, typically proprietary ones like GPT-4. Previous studies yield mixed results: Khare et al.~\cite{khare2023understanding} showed that LLMs perform well on \emph{synthetic datasets} but not promising on real-world datasets.
Experimentation with different prompting strategies, notably variations of Chain-of-Thought (CoT) prompts, has further shown promising results~\cite{ullah2023can}. 
Integrating LLMs into larger frameworks has shown promise in detecting specific vulnerabilities, such as Use Before Initialization vulnerabilities~\cite{li2024enhancing} and smart contract vulnerabilities~\cite{sun2024llm4vuln}.
We analyzed many of these Code LM-based \vd methods in this paper and showed that in a very realistic setting none of them performs well.

\paragraph*{Empirical Analysis of Deep-learning-based Vulnerability Prediction (DLVP)}
Several works~\cite{chakraborty2021deep,steenhoek2023empirical,steenhoek2023language,risse2023limits,micelibarone_larger_2023} have pointed out that while DLVP models make correct predictions, they do so for the wrong reasons; 
they often rely on ``spurious features'' that are not the root cause of the vulnerabilities. Chen et al.~\cite{chen2023diversevul} find that, through a large-scale evaluation involving 26k vulnerable functions across 300 projects and 150 CWEs, DLVP lacks generalization to unseen projects and is still far from being deployed in the industry. Some of the prior works~\cite{risse2023limits,ullah_can_2023} focus on the lack of robustness of ML-based vulnerability detection algorithms against semantically preserving modifications. Another recent work~\cite{steenhoek2023language} attempts to measure how much models pick up the bug semantics through interpretability techniques involving the attention mechanism and shows that extra annotation on the bug semantics also improves the model's performance. Our current work complements these lines of work by focusing more on benchmark creation and evaluation techniques.

\vspace{1mm}
\section{Conclusions}
In this paper, we uncover significant limitations in existing vulnerability detection datasets, such as poor data quality, low label accuracy, and high data duplication rates, as well as the limited practical utilities of current evaluation metrics.

In response to these concerns, we present \dataset, accompanied by updated evaluation criteria designed to more accurately gauge practical effectiveness of Code LM-based vulnerability detectors. Through a series of experiments on \dataset, we find that even with efforts to improve performance using sophisticated methods and expansive models, existing Code LMs consistently fail to meet the demands of effective vulnerability detection in practical settings. This underscores the urgent requirement for fundamentally innovative approaches in training Code LMs for security applications, while also establishing a new benchmark for evaluating their efficacy.
\section*{Acknowledgement}

This material is based upon work supported by the National Science Foundation under grants 2229876, 2154873, 2221943, 2313055, 1845893, and 2107405, and by the Department of Homeland Security, IBM, the Center for AI Safety Compute Cluster, the Noyce Foundation, C3.ai DTI, and the KACST-UCB Center of Excellence for Secure Computing.  Any opinions, findings, conclusions, or recommendations expressed in this material are those of the author(s) and do not necessarily reflect the views of the sponsors.

\small
\bibliographystyle{unsrt}
\bibliography{main}

\clearpage
\section*{Supplementary Material}

\begin{table*}[h]
\caption{Detailed breakdown for label error analysis. The error categories are adopted from Chen et al.~\cite{chen2023diversevul}.}
	\centering
	\begin{tabular}{l | c | c  c  c }
		\toprule
		\multirow{3}{*}{\textbf{Benchmark}} & \multirow{3}{*}{\textbf{Correct Label}} & \multicolumn{3}{c}{\textbf{Wrong Label}} \\
        & & \multirow{2}{*}{\begin{tabular}{@{}c@{}}{Vulnerability Spread}\\{Across Multiple Functions}\end{tabular}} & \multirow{2}{*}{Relevant Consistency} & \multirow{2}{*}{Irrelevant} \\
        & & & & \\
        \midrule
        \textbf{SVEN~\cite{he2023sven}} & 94\% & 0\% & 0\% & 6\% \\
        \textbf{CodeXGLUE~\cite{lu2021codexglue, zhou2019devign}} & 24\% & 18\% & 0\% & 58\% \\
        \textbf{VulnPatchPairs~\cite{risse2023limits}} & 36\% & 10\% & 14\% & 40\% \\
        \midrule
        \textbf{\dataset-\textsc{OneFunc}} & 86\% & 4\% & 4\% & 6\%\\
        \textbf{\dataset-\textsc{NVDCheck}} & 92\% & 4\% & 2\% & 2\%\\
        \bottomrule
	\end{tabular} 
	
	\label{tab:label_accuracy_breakdown}
\end{table*}

\subsection{Detailed Label Error Analysis}
As demonstrated in Section II.B and Section III.B of the main paper, we assess 50 vulnerable functions randomly sampled from each dataset, including CodeXGLUE~\cite{lu2021codexglue, zhou2019devign}, VulnPatchPairs~\cite{risse2023limits} and SVEN~\cite{he2023sven}, along with vulnerable functions identified by our two techniques, \dataset-\textsc{OneFunc} and \dataset-\textsc{NVDCheck}. The manual analysis results are shown in Table 1 of the main paper. In this section, we provide a more detailed analysis of the wrongly labeled functions. 

Chen et al.~\cite{chen2023diversevul} identify three primary sources of labeling errors. First, irrelevant functions—those subject to formatting, non-functional alterations not linked to security fixes, or contained within commits unrelated to security issues—were incorrectly tagged as vulnerable. Second, the vulnerability may spread across multiple functions, which does not match our goal of training neural networks to learn whether a single function is vulnerable. Third, benign functions undergoing modifications during vulnerability fixes, such as parameter list adjustments for consistency with altered vulnerable functions, were erroneously marked as vulnerable. Building on these three categories of labeling errors, we dissect the inaccuracies identified in our manual analysis, with the detailed breakdown presented in Table~\ref{tab:label_accuracy_breakdown}.

Our manual analysis reveals that only 24\% of the 50 data points evaluated from CodeXGLUE and 36\% from VulnPatchPairs are genuinely vulnerable. This is noteworthy considering CodeXGLUE's manual efforts in labeling security-related commits and VulnPatchPairs is built on CodeXGLUE. Contrasting our findings, Croft et al.~\cite{croft2023data} acknowledge a labeling accuracy issue with CodeXGLUE, yet their review finds 80\% of their sampled vulnerable functions correctly labeled. This divergence largely stems from our more stringent criteria for identifying vulnerable functions.

First, after a more careful examination, we find that 58\% of the samples from CodeXGLUE and 40\% from VulnPatchPairs are irrelevant to security and most of them originate from commits unrelated to security issues. Examples include commits mentioning ``instrument memory management''\footnote{\url{https://github.com/qemu/qemu/commit/cd245a1}} or ``remove ad-hoc leak checking code''.\footnote{\url{https://github.com/qemu/qemu/commit/7d1b009}} Previous human annotators may misinterpret these commits as security-related due to certain keywords in the commit messages, despite their irrelevance to actual security issues. Additionally, we discover examples like commits focused solely on code migration\footnote{\url{https://github.com/qemu/qemu/commit/60fe637}} or operating system compatibility improvements\footnote{\url{https://github.com/qemu/qemu/commit/b981289}}, which are not security-related at all.

Second, contrary to the approach taken by Croft et al.~\cite{croft2023data}, our methodology labels a function as vulnerable only if it independently constitutes a security risk. For instance, addressing a race condition, as seen in certain commits\footnote{\url{https://github.com/qemu/qemu/commit/c5a49c6}}, demands a comprehensive understanding of the system architecture, making it improper to assess a function's vulnerability in isolation. Previous annotators often mark all functions associated with such race conditions as vulnerable. Another case\footnote{\url{https://github.com/qemu/qemu/commit/eefa3d8}} involves a denial-of-service (DoS) vulnerability linked to the repetitive invocation of the \texttt{recvmsg} function. The \texttt{qio\_channel\_websock\_encode} function, which merely shifts some values, does not directly lead to a DoS threat. Thus, we categorize \texttt{qio\_channel\_websock\_encode} as non-vulnerable, diverging from prior analyses. 

Third, Croft et al.~\cite{croft2023data} label the callers of vulnerable functions as vulnerable. Conversely, we classify such functions as benign because they are altered solely to align with security updates. One case\footnote{\url{https://github.com/FFmpeg/FFmpeg/commit/073c259}} is the \texttt{wma\_decode\_init} function that invokes a vulnerable function \texttt{init\_vlc}. In our analysis, the function \texttt{wma\_decode\_init} is benign.

\vspace{-2mm}
\begin{figure}[ht]
    \centering
    \includegraphics[width=0.95\linewidth]
    {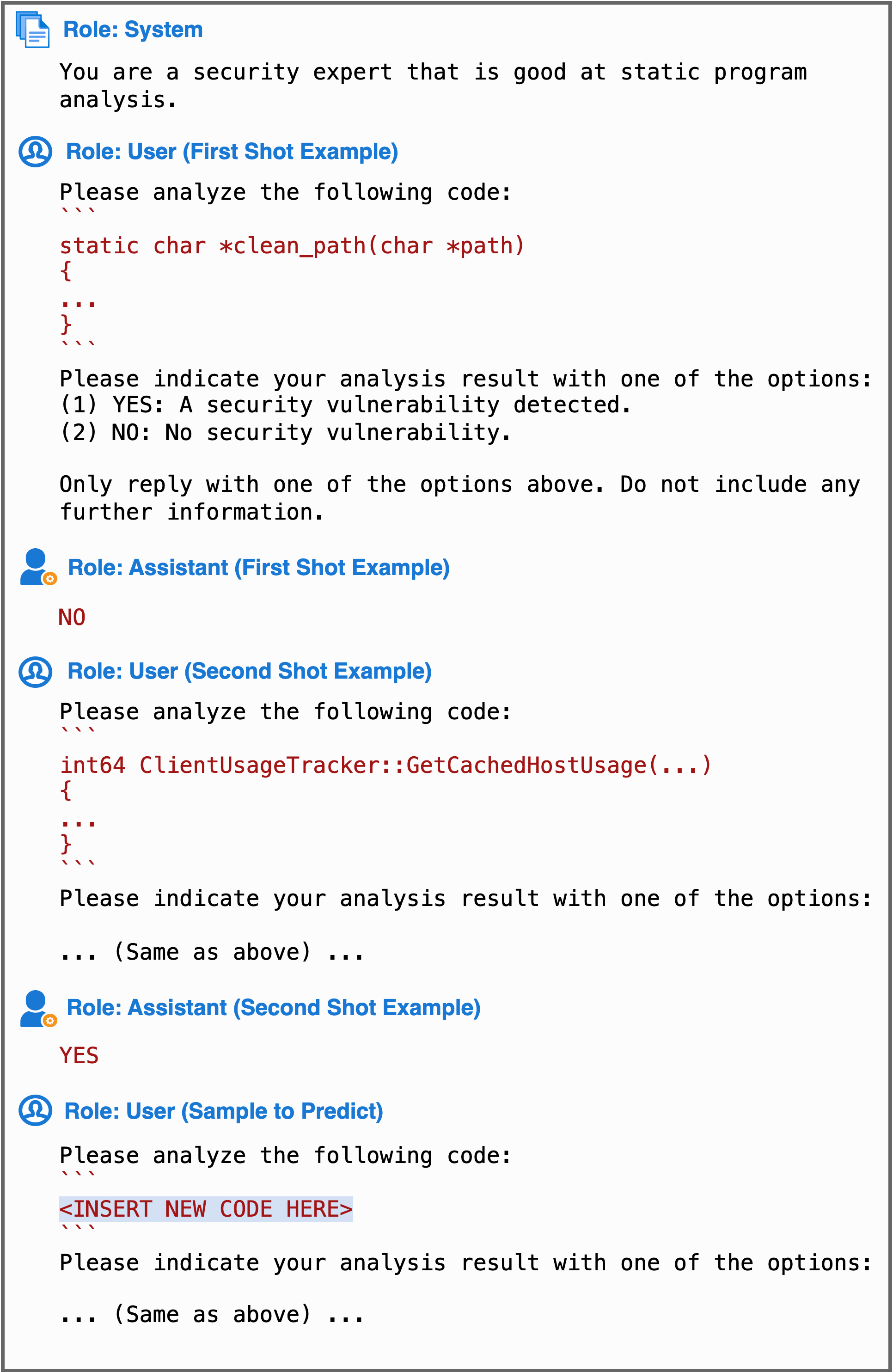}
    \caption{The template for the two-shot prompt.}
    \label{fig:two_shot_prompt}
\end{figure}
\begin{figure}[h]
    \centering
    \includegraphics[width=0.95\linewidth]
    {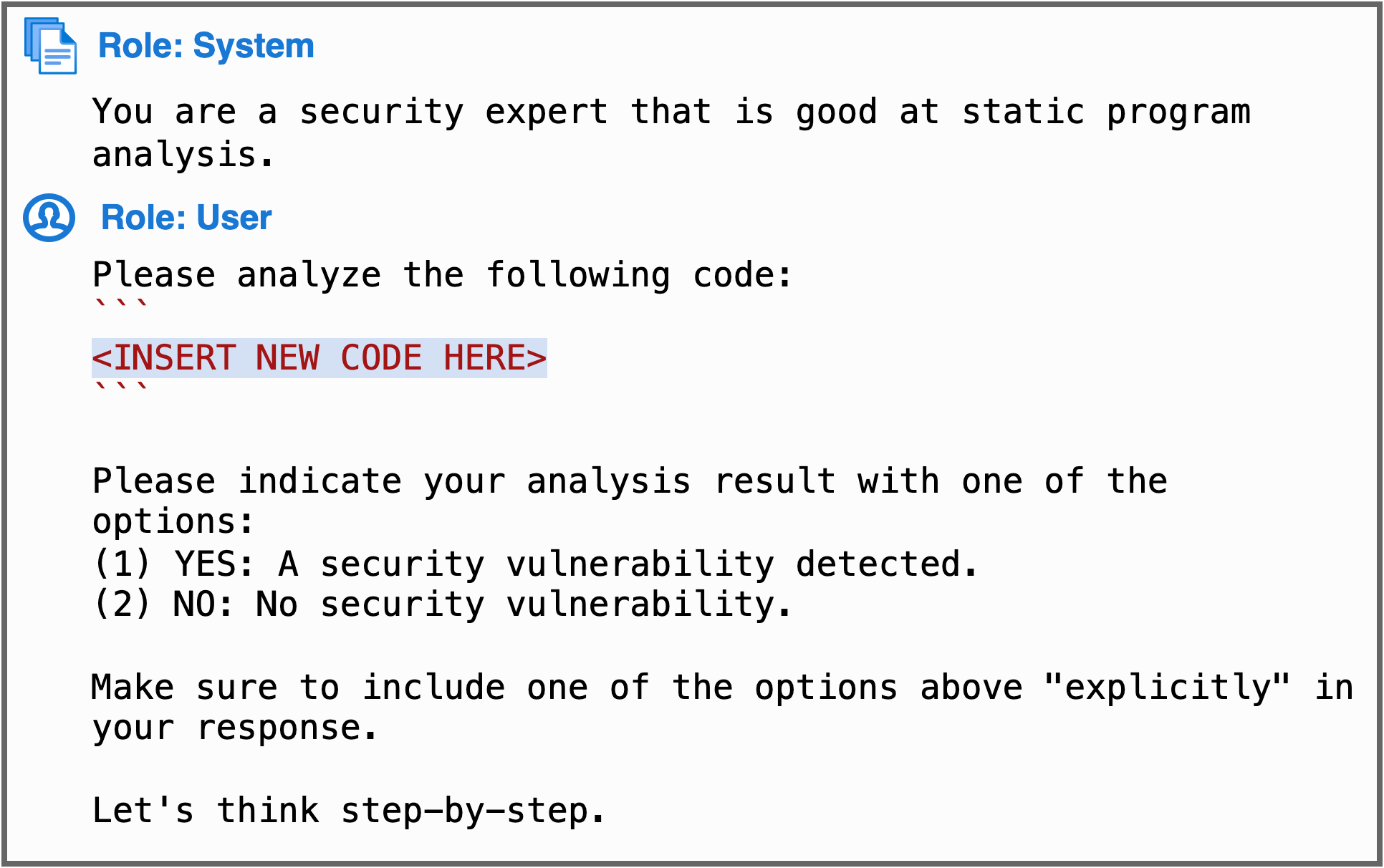}
    \caption{The template for the chain-of-thought prompt}
    \label{fig:cot_prompt}
\end{figure}

\subsection{Prompts for Open AI Models}

In this section, we show the template we used for prompting the OpenAI models. 

Figure~\ref{fig:two_shot_prompt} shows the template of two-shot prompting, where we start from a system prompt to warm up the model with the task it will deal with. Then it is followed by one benign example and one positive example. In the end, the new code to be predicted will be wrapped into the position \texttt{<INSERT\_NEW\_CODE\_HERE>}.

Figure~\ref{fig:cot_prompt} shows the template of chain-of-though reasoning for detecting the vulnerability. Similarly, the new code to be predicted will be wrapped into the position \texttt{<INSERT\_NEW\_CODE\_HERE>}.

\end{document}